\documentclass[twocolumn,showpacs,preprintnumbers,amsmath,amssymb]{revtex4}

\usepackage{graphicx}
\usepackage{dcolumn}
\usepackage{bm}
\usepackage{epsf}

\newcommand{\bea}{\begin{eqnarray}}
\newcommand{\eea}{\end{eqnarray}}
\newcommand{\eq}[1]{eq.~(\ref{#1})}

\newcommand{\Eqs}[2]{Eqs.(\ref{#1},\ref{#2})}

\newcommand{\ur}[1]{(\ref{#1})}

\newcommand{\beq}{\begin{equation}}
\newcommand{\eeq}{\end{equation}}
\newcommand{\la}[1]{\label{#1}}
\newcommand{\ba}{\begin{array}}
\newcommand{\ea}{\end{array}}

\newcommand{\half}{{\textstyle{\frac{1}{2}}}}
\newcommand{\at}{\overline{10}}

\newcommand{\nn}{\nonumber}

\begin{document}

\title{Comment on the $\Theta^+$ width and mass}
\author{Dmitri Diakonov$^{a,b,c}$}
\author{Victor Petrov$^c$}
\author{Maxim Polyakov$^{c,d}$}
\vskip 0.3true cm

\affiliation{
$^a$ Thomas Jefferson National Accelerator Facility, Newport News, VA 23606, USA\\
$^b$ NORDITA,  Blegdamsvej 17, DK-2100 Copenhagen, Denmark\\
$^c$ St. Petersburg Nuclear Physics Institute, Gatchina, 188 300, St. Petersburg, Russia\\
$^d$ Universit\'e de Liege, B-4000 Liege 1, Belgium
}

\date{May 18, 2004}

\begin{abstract}
We discuss the relatively low mass and narrow width prediction for 
the exotic baryon $\Theta^+$, and comment on recent statements 
by R.L. Jaffe on the subject. We reaffirm that a narrow width
of $3.6-11.2\,{\rm MeV}$ follows from the equations of our 1997 paper.
\end{abstract}

           

\maketitle

In the 1997 paper~\cite{DPP1997} we predicted a relatively light 
($M\!\approx\!1530\,{\rm MeV}$) and narrow ($\Gamma\!\leq\!15\,{\rm MeV}$) exotic 
baryon with strangeness $+\!1$, isospin zero and spin-parity $\half^+$. 
The paper was published at a time when all previous searches of exotic baryons 
for thirty years were in vain~\cite{PDG1986}, and the latest phase shift 
analysis~\cite{HARW1992} summarizing the $KN$ scattering data showed no 
signs of a resonance in this energy range. The prediction motivated
and oriented new experimental searches, and in the end of 2002 the first 
independent observations of the exotic baryon in $\gamma{\rm C}$ 
\cite{Osaka} and $K^+{\rm Xe}$ \cite{ITEP} reactions were announced, 
followed by important confirmation in about ten experiments by spring 
2004~\cite{JLab,ELSA,neutrino,HERMES,Protvino,Juelich,Dubna,ZEUS}. 
In one year, $\Theta^+$, as it has been named following our suggestion~\cite{Letter},
made it to the Review of Particle Properties~\cite{PDG2004}. At the same
time, there have been several experiments where $\Theta^+$ has not been 
seen~\cite{no-Theta-1,no-Theta-2}. Therefore, one is now looking forward 
to the next tour of dedicated experiments with higher statistics, for the 
issue to be finally resolved.  

Not unnaturally, theoretical \cite{Cohen,DPlargeNc,Klebanov,Pobylitsa,Praszalowicz}, 
phenomenological ~\cite{EKP,Schweitzer} and arithmetical~\cite{Jaffe} aspects of 
the 1997 paper have been recently scrutinized. In the latter comment, Jaffe 
states that we have made an arithmetic mistake in Ref.~\cite{DPP1997}
and that, being corrected, it would lead to a prediction of $30\,{\rm MeV}$ for the
$\Theta$'s width. Although the questions of the $\Theta$'s mass and width have been 
already discussed in great detail by Ellis, Karliner and Praszalowicz~\cite{EKP} 
who have basically confirmed the calculations of Ref.~\cite{DPP1997} and explored
the unbiased theoretical and phenomenological uncertainties, we feel it necessary to 
respond directly to Jaffe's comment. To that end, we have first of all to remind
briefly the calculation of the $\Theta$ width.

The equation (56) for $\Theta$'s width from Ref.~\cite{DPP1997} reads:       
\beq
\Gamma_\Theta = \frac{3\,G_{\at}^2}{2 \pi[m_N+m_\Theta]^2}
\frac{m_N}{m_\Theta}\,\frac{1}{5}\,|{\bf p}|^3
\la{Gamma_Theta}\eeq
where ${\bf p}$ is the kaon momentum in the decay $\Theta^+\!\to\!NK$, 
and we have neglected the small correction due to the antidecuplet-octet mixing. The
pseudoscalar antidecuplet-octet transition constant $G_{\at}$ is expressed through 
the more fundamental symmetry constants $G_{0,1,2}$ (Table 2 of Ref.\cite{DPP1997}): 
\beq
G_{\at}=G_0-G_1-\frac{1}{2}G_2.
\la{Gat}\eeq
The constant $G_2$ is small as it is related to the singlet nucleon axial constant 
$g_A^{(0)}$, better known as the fraction of nucleon spin carried by quarks' spin 
(equation (53) of~\cite{DPP1997}):
\beq
G_2=\frac{2m_N}{3F_\pi}\,g_A^{(0)}\simeq 2
\la{G2}\eeq 
if $g_A^{(0)}\simeq 0.3\pm 0.1$ is used~\cite{FJ}. Another combination of $G_{0,1,2}$ 
determines the standard pion-nucleon coupling constant (equation (50) of~\cite{DPP1997}):
\beq
g_{\pi NN}=\frac{7}{10}\left(G_0+\frac{1}{2}G_1+\frac{1}{14}G_2\right)\simeq 13.3
\la{gpiNN}\eeq
where we have substituted the present value of the pseudoscalar pion-nucleon 
constant~\cite{ELT}. \Eqs{G2}{gpiNN} allow one to find the combination
\beq
G_{10}=G_0+\frac{1}{2}G_1\simeq 18.9
\la{G10}\eeq
from phenomenology, but not $G_{\at}$ determining the $\Theta$ decay. 
Therefore, in Ref.~\cite{DPP1997} we have used an additional theoretical input,
\beq
\frac{G_1}{G_0}=\rho
\la{ratio}\eeq 
with $\rho$ ranging from 0.4 to 0.6 as it follows from the estimates in the chiral
quark soliton model~\cite{G1}. \Eqs{G10}{ratio} fix the $G_{\at}$ constant
needed to compute the $\Theta$ width. In Ref.~\cite{DPP1997} we have used the following 
values of the masses: $m_\Theta=1530\,{\rm MeV},\,m_N=940\,{\rm MeV},\,m_K=495\,{\rm MeV}$ 
leading to the kaon momentum in the decay
\bea\nn
|{\bf p}|&=&\frac{\sqrt{m_\Theta^4\!+\!m_N^4\!+\!m_K^4\!-\!2m_\Theta^2m_N^2\!
-\!2m_\Theta^2m_K^2\!-\!2m_N^2m_K^2}}{2\,m_\Theta}\\
&=&254\,{\rm MeV}.
\la{pK}\eea
Putting these numbers into \eq{Gamma_Theta} one gets finally
\beq
\Gamma_\Theta=\left\{\begin{array}{cc}
3.6\,{\rm MeV}, & \rho=0.6,\\
6.7\,{\rm MeV}, & \rho=0.5,\\
11.2\,{\rm MeV}, & \rho=0.4. \end{array}\right.
\la{Gamma_values}\eeq 
In view of the theoretical uncertainty in the estimate of the $G_1/G_0$ ratio, we have
concluded that the exotic baryon width must be ``less than $15\,{\rm MeV}$'', and put
it in the Abstract in Ref.~\cite{DPP1997}. In the original paper we have neglected 
the small $G_2$ in the final estimate \ur{Gamma_values} \footnote{In 1997 the fraction 
of the proton spin carried by quark spins was compatible with zero.} but took into account 
the small correction from antidecuplet-octet mixing and hence obtained a slightly 
higher upper limit, $\Gamma_\Theta\leq 15\,{\rm MeV}$ at $\rho=0.4$. We noted however 
(before eq.(56)) that it was a conservative estimate and that $\Theta$ could be more narrow. 
The key element in the narrow width prediction is the strong cancellation 
between the $G_0,G_1$ and $G_2$ contributions, which we have noticed. Furthermore, 
we have noticed that in the non-relativistic limit implying $G_1/G_0=4/5,\,G_2/G_0=2/5$ 
one gets zero $G_{\at}$ and hence zero $\Theta$ width. 

A separate issue is the widths of the usual $\left({\bf 10},\frac{3}{2}\right)$ baryons.
They are not directly related to the above estimate of the $\Theta$ width, however their
discussion has been included in Ref.~\cite{DPP1997} for completeness. 
For spin 1/2 decays, there is only one formfactor involved, and in whatever way one treats
kinematical factors in the case of $\left({\bf 8},\frac{1}{2}\right)$ baryons, one can repeat 
the same for the $\left({\bf \at},\frac{1}{2}\right)$ ones. For spin 3/2 decays, it becomes 
more ambiguous. In the academic limit of large number of colors $N_c$, the decuplet-to-octet 
transition constant $G_{10}$ is determined by \eq{G10} and thus related to the octet 
$g_{\pi NN}$ pseudoscalar coupling. However, in reality an additional uncertainty arises for spin 
$\frac{3}{2}\to\frac{1}{2}$ decay widths: Should one use Adler's formfactors~\cite{Adler}, 
treat the spin 3/2 in the relativistic Rarita--Schwinger formalism and take the exact 
spin 3/2 density matrix to compute the phase volume, or should one rather estimate the 
transition matrix element by non-relativistic formulae and then simply multiply it by 
the relativistic phase volume? Should the symmetry relation \ur{G10} be imposed on the 
axial vector constants or rather on the pseudoscalar ones? Depending on the choice one makes, 
one gets different functions of the mass ratio $m_1/m_2$ in the expressions
for the spin 3/2 widths ($m_1$ is the initial and $m_2$ is the final baryon mass in the decay). 
This mass ratio is unity in the large $N_c$ limit, since in this limit both $N$ and $\Delta$ 
are infinitely heavy non-relativistic particles, such that it does not matter which way one 
decides to resolve the ambiguities, but in the real world it does matter since $\Delta$ is 
30\% heavier than the nucleon. This ambiguity is encountered by all people who have attempted 
to fit the decuplet decays from the knowledge of the $g_{\pi NN}$ constant, be it from large-$N_c$ 
or non-relativistic quark considerations~\cite{HHM}. There are infinitely many ways how 
one can resolve this ambiguity, and any of them is guess work from the strict theory 
point of view as it corresponds to some particular hypothesis how to sum up an infinite 
series of unknown corrections in quark masses and $1/N_c$. It reflects the true 
situation and the actual theoretical accuracy with which one is able to compute 
the spin 3/2 widths from the large-$N_c$ considerations.

When working on the 1997 paper, we have noticed that if, for the spin 3/2 decays, 
one rescales the symmetry relation \ur{G10}, by the $m_1/m_2$ ratio, the {\em four} 
known decuplet decay rates are described very satisfactory and are in accordance 
with the value of the $g_{\pi NN}$ constant. Indeed, with this rescaling eqs.(42-45) 
of Ref.~\cite{DPP1997} should read:
\bea
\nn
\Gamma(\Delta \to N\pi)&=&
\frac{3 G_{10}^{2}}{2\pi (m_\Delta+m_N)^2} |{\bf p}|^3
\frac{m_\Delta}{m_N}\cdot\frac{1}{5}\\
\nn
&=& 110\,{\rm MeV}\;{\rm vs}\;100\!-\!125\,{\rm MeV}\;({\rm exp.}),\\
\nn
\Gamma(\Sigma^* \to \Lambda\pi)&=&
\frac{3 G_{10}^{2}}{2\pi (m_{\Sigma^*}+m_\Lambda)^2}
|{\bf p}|^3 \frac{m_{\Sigma^*}}{m_\Lambda}\cdot\frac{1}{10}\\
\nn
&=& 30.5\,{\rm MeV}\;{\rm vs}\;32.6\,{\rm MeV}\;({\rm exp.}),\\
\nn
\Gamma(\Sigma^* \to \Sigma\pi)&=&
\frac{3 G_{10}^{2}}{2\pi (m_{\Sigma^*}+m_\Sigma)^2}
|{\bf p}|^3 \frac{m_{\Sigma^*}}{m_\Sigma}\cdot\frac{1}{15}\\
\nn
&=& 3.7\,{\rm MeV}\;{\rm vs}\;4.4\,{\rm MeV}\;({\rm exp.}),\\ 
\nn
\Gamma(\Xi^* \to \Xi\pi)&=&
\frac{3 G_{10}^{2}}{2\pi (m_{\Xi^*}+m_\Xi)^2} |{\bf p}|^3
\frac{m_{\Xi^*}}{m_\Xi}\cdot\frac{1}{10}\\
&=& 8.8\,{\rm MeV}\;{\rm vs}\;9.3\,{\rm MeV}\;({\rm exp.}),
\la{decuplet_decay}\eea
where ${\bf p}$ is the pion momentum and $G_{10}\simeq 18.9$ from \eq{G10} is 
used~\footnote{In the numerics, we have averaged the masses over the isospin components,  
and took the experimental widths from the PDG-2002 edition, which explains a small deviation
in numbers from eqs.(42-45) of the earlier Ref.~\cite{DPP1997}.}. It is interesting 
that if one considers only non-strange baryons, the large-$N_c$ relation between the $\Delta$
width and the $g_{\pi NN}$ constant for two flavors is well satisfied without the rescaling 
of the spin-3/2 constant by the mass ratio~\cite{ANW}. It shows once more that there is some 
arbitrariness in the theoretical treatment of the strange quark mass and $1/N_c$ corrections 
to the decuplet decays. Unfortunately, in the write-up a year later after the actual calculations, 
we did not discuss the problem of the spin-3/2 decays (which was anyhow secondary to the 
more important issues related to the suggested new spin-1/2 antidecuplet of baryons) 
and wrote all equations universally as if they were for spin-1/2 decays, 
but left the numerical values of the widths in the right hand sides computed from the 
rescaled formulae. Weigel discovered this inconsistency~\cite{Weigel} 
and communicated it to one of us (M.P.) who acknowledged the mistake. We apologize to 
those who might have been lead into confusion. However, we told very many people 
about this mistake, including the authors of Ref.~\cite{EKP} and Jaffe. 

In his comment~\cite{Jaffe}, Jaffe suggests that one has to take our mistake in the analytical 
expressions for the decuplet decays at face value, fit the $\Delta$ decay with a larger value 
of $G_{10}\simeq 25$, get an {\em unacceptably large} $g_{\pi NN}\simeq 17.5$ from \eq{gpiNN} and 
correspondingly a larger value of the $\Theta$ width. However, this is not the way to proceed. 
If the $g_{\pi NN}$ constant matches the spin-3/2 decuplet decays, one can use {\em either} 
$g_{\pi NN}$ {\em or} the decuplet widths as an input to estimate the width of $\Theta^+$, since 
it gives the same. This is the case when one rescales $G_{10}$ by the $m_1/m_2$ mass ratio, 
as in eqs. ({\ref{decuplet_decay}}), or does not rescale the constant but uses the
2-flavor relation between $g_{\pi NN}$ and $g_{\pi N\Delta}$~\cite{ANW}. 
If one does not succeed to match them, one uses the phenomenological value of the spin-1/2 
$g_{\pi NN}$ constant to get the same small width of the spin-1/2 $\Theta^+$ as described above, 
but faces an unrelated problem how to explain the theoretically more dubious spin-3/2 decays. 

Let us emphasize it again: The narrow width prediction for the $\Theta^+$ can be
obtained without even mentioning the decuplet. It is founded on the dynamical cancellation 
in the $G_{\at}$ constant, see above~\footnote{Recently Praszalowicz has demonstrated that 
this cancellation persists at any number of colors~\cite{Praszalowicz}.}. In fact the present 
day theoretical uncertainty in how ``deep'' is this cancellation, resulting in the spread 
in \eq{Gamma_values}, is greater than the theoretical uncertainty in the decuplet 
decays~\footnote{It should be mentioned that a more careful analysis of the semileptonic 
hyperon decays in Ref.~\cite{Rathke} lead to the estimate $\Gamma_\Theta<5\,{\rm MeV}$.}.  

We would like to comment on two other statements by Jaffe~\cite{Jaffe}. Both 
comments are historic but elucidate physics as well. 20 years ago, in the 
Fall of 1983 a seminal paper by Witten appeared~\cite{Witten}, where there was
a brief Note Added in Proof with the now famous quantization condition 
that only those $SU(3)$ baryon multiplets appear as rotational states of a chiral 
soliton in the 3-flavor space, which have hypercharge $Y=N_c/3=1$, and that the 
spin of the allowed multiplet is related to the number of particles with that hypercharge. 
Since no derivation was given, several groups~\cite{SU3} derived this result in 1984-85 in 
their own manner~\footnote{From this list, Guadagnini, Mazur et al. and Jain and Wadia
quantized the $SU(3)$ skyrmion independently; other authors cite Guadagnini's 
paper.}, including two of the present authors~\cite{DP_ITEP}. 
In February 1984 one of us (D.D.) gave lectures on this particular subject at the ITEP 
Winter School, where the quantization of ordinary and SU(3) flavor rotations of a chiral 
soliton has been exhaustively explained and its implications for higher rotational 
states discussed. In the published version of those lectures D.D. and V.P. have written on p. 90: 
``We thus come to the conclusion that the lowest states of the chiral soliton 
are the octet with spin 1/2 and the decuplet with spin 3/2. We leave it as an exercise 
to the reader to find other multiplets which can be interpreted as rotational
states of a spherically symmetric chiral soliton." When one knows Witten's quantization 
conditions, one opens any book on $SU(3)$ and observes that the next baryon multiplets 
satisfying them are $\left({\bf \at},\frac{1}{2}\right)$, $\left({\bf 27},\frac{3}{2}\right)$, 
$\left({\bf 27},\frac{1}{2}\right)$ and so on. In 1984 the potential existence of an exotic 
{\it anti\-decuplet} of baryons became plain to the skyrmion community. 

Knowing about the potential existence of the baryon antidecuplet, many people
including ourselves made estimates of its masses as early as 1984 with whatever
calculational tools available at that time. In his comment, Jaffe gives credit to
Praszalowicz for predicting the correct mass of the $\Theta$. Indeed, Praszalowicz
mentioned the mass of the lowest baryon, the future $\Theta$, at 1530 MeV~\cite{Krakow}
as following from the Skyrme model. However, it may be worth pointing out 
that there were inconsistencies in that calculation. First of all, in the unrealistic 
Skyrme model one has to make a difficult choice between having the nucleon and 
$\Delta$ masses correct and the $F_\pi$ constant completely wrong
($F_\pi\simeq 64\,{\rm MeV}\;{\rm vs}\; 93\,{\rm MeV}\;({\rm exp.})$) \cite{ANW,Krakow},  
or {\it vice versa}. To make matters worse, in the Skyrme model the normal term proportional
to strangeness in the baryon splitting, is absent. Therefore, to account approximately
for the observable splitting in strangeness in the octet and decuplet baryons, one 
has to employ the Yabu--Ando method of involving higher corrections in the strange 
quark mass, which adds new free parameters. As seen from Fig.~1 in Ref.~\cite{Krakow} 
the choice of a free parameter corresponding to $m_\Theta=1530\,{\rm MeV}$, 
corresponds also to poor masses of the normal $N,\Lambda$ and $\Sigma$. If one makes 
a better fit to the known baryons, $\Theta$ shifts to 1340 MeV, i.e. below the threshold 
for strong decays. The $F_\pi$ constant remains 1.5 times less than it should be. If anything, 
in this very useful paper Praszalowicz demonstrated that the Skyrme model was unfit 
to make accurate predictions. As the author correctly noted himself, 
``one has to express criticism against [the Yabu--Ando method], as it sums up an arbitrary 
subseries of the strange quark masses, neglecting other terms of the same order"~\cite{Siegen}. 
The same remark concerns the estimate of $m_\Theta$ by Walliser who, using basically 
the same Skyrme model but another version of the $SU(3)$ symmetry breaking, obtained it at about 
1700 MeV~\cite{Walliser}. In another variant of the Skyrme model Walliser got a remarkable
$m_\Theta=1550\,{\rm MeV}$! The earliest printed estimate we found was in the 1984 paper by  
Biedenharn and Dothan~\cite{BD} who evaluated the antidecuplet mass (without splitting)
at $m_N+600\,{\rm MeV}=1540\,{\rm MeV}$, with the conclusion that it was an artifact of the model.
In short, one could get various ``predictions" for the $\Theta$ mass from the
Skyrme model, depending on what observables for the established hadrons one was prepared 
to sacrifice. Since there were too many inherent inconsistencies inside the model,
none of these authors seemed to have taken the antidecuplet for real. 

As to the widths of the antidecuplet baryons, in the standard version of the Skyrme
model the constants $G_{1,2}$ discussed above are zero. There is no
possibility for the dynamical cancellation of the $G_0$ constant, leading to the narrow 
$\Theta$ width, which therefore could have never been obtained in the Skyrme model.  

It was not until the modern chiral quark soliton model of baryons~\cite{DP_CQSM} 
has been developed that one could estimate the masses and the widths of 
the antidecuplet baryons in a consistent way. The instanton-based 
chiral quark model gives a coherent picture of mesons and baryons; in particular
it explains the basic facts about baryons, which are mysterious in the conventional 
constituent quark models: why quark spins carry only 1/3 of the nucleon spin~\cite{WY}, 
why the nucleon sigma term is 4 times larger than counted from quarks~\cite{DPPrasz}, 
why there are many antiquarks in the nucleon at a low virtuality~\cite{SF}, 
why the sea antiquarks are flavor-asymmetric~\cite{FA}, and many other features, 
not to mention an overall fair description of masses, magnetic moments and 
formfactors~\cite{Review}. Simultaneously, it gives the reason why constituent quark models
are in many cases successful. There are basically no free parameters in the 
model as it follows from the QCD lagrangian~\cite{DP_Ioffe}. When one feels 
that the known basic facts are understood, one may risk to make a prediction, 
despite a heavy pressure from the unsuccessful attempts to find exotic baryons in the past.
 
The accurate prediction of the $\Theta$ mass in Ref.~\cite{DPP1997} was to some extent 
a luck. It was in part based on the use of a certain value of the nucleon sigma term
resulting in a large splitting in the antidecuplet, and on identifying the $N(1710)$ 
resonance as the antidecuplet member. These were legitimate assumptions in 1997 
but later experimental data shifted the sigma term to larger values~\cite{PASW,Schweitzer},
resulting in a smaller antidecuplet splitting~\cite{DP_mixing}, while a new 
analysis~\cite{AAPSW} indicated that the former $N(1710)$ might be in fact 
lighter~\footnote{Preliminary data from the $\gamma n\to \eta n$ reaction at GRAAL 
in Grenoble indicate a narrow $N^*$ resonance at $1670\,{\rm MeV}$ whose properties 
seem to fit its antidecuplet nature~\cite{Kouznetsov}.}. It seems like there
is a lucky cancellation between the inaccuracies in the two inputs, each on the scale 
of a few tens MeV. 

The important points of Ref.~\cite{DPP1997} were a)~that the exotic baryon $\Theta^+$
with spin-parity $\half^+$ must exist, b)~that it must be relatively light, 
c)~that it must be narrow. These points came from the experience in the quantitative 
description of the properties of the usual hadrons. The relatively low mass and width of 
the antidecuplet are explained in qualitative terms in Ref.~\cite{DP_mixing}.  

We thank K. Goeke, R. Jaffe, K. Hicks, B. Holstein, M. Karliner, M. Praszalowicz and 
M. Strikman for useful comments on the first version.

\end{document}